\documentclass[twocolumn,aps,prl,amsmath,amssymb]{revtex4-1}
\usepackage[latin9]{inputenc}
\usepackage{mathrsfs}
\usepackage{amsmath}
\usepackage{amssymb}
\usepackage{graphicx}
\usepackage{esint}

\makeatletter

\usepackage{epsfig}
\usepackage{bm}
\usepackage{dcolumn}
\usepackage{color}

\makeatother

\begin{document}
\title{Two-dimensional exciton-polariton interactions beyond the Born approximation}
\author{Hui Hu$^{1}$, Hui Deng$^{2,3}$, and Xia-Ji Liu$^{1}$}
\affiliation{$^{1}$Centre for Quantum Technology Theory, Swinburne University
of Technology, Melbourne, Victoria 3122, Australia}
\affiliation{$^{2}$Department of Physics, University of Michigan, Ann Arbor, MI
48109, USA}
\affiliation{$^{3}$Applied Physics Program, University of Michigan, Ann Arbor,
MI 48109, USA}
\date{\today}
\begin{abstract}
We provide a many-body theory for the interactions of two-dimensional
excitons and polaritons beyond the Born approximation. Taking into
account Gaussian quantum fluctuations via the Bogoliubov theory, we
find that the two-body interaction strength in two-dimensions has
an inverse logarithmic dependence on the scattering length and ground
state energy. This leads to a vanishing exciton interaction strength
in the zero-momentum limit but a finite polariton interaction strength
due to strong light-matter coupling. We also derive the exact Tan
relations for exciton-polaritons and calculate Tan's contact coefficient.
We show the polariton interaction strength and Tan's contact both
exhibit an anomalous enhancement at red photon-exciton detuning when
the scattering length is large. Our predictions may provide a qualitatively
correct guide for studies of exciton and polariton nonlinearities,
and suggest a route to achieving strongly nonlinear polariton gases. 
\end{abstract}
\maketitle
Exciton-polaritons are elementary excitations of a semiconductor formed
via strong coupling between excitons and photons \cite{weisbuch_observation_1992}.
Due to their half-matter, half-light nature, they form a unique platform
for a wide range of novel nonlinear phenomena that are absent in linear
optical systems and hard to access in pure matter systems \citep{keeling_collective_2007,Deng2010,Carusotto2013,Byrnes2014,Fraser2016,Sanvitto2016},
ranging from a variety of many-body quantum phases \cite{Deng2002,roumpos_power-law_2012,Ravets2018},
resonant parametric scattering \cite{savvidis_angle-resonant_2000,baumberg_parametric_2000},
ultra-low threshold lasing \cite{deng_polariton_2003,schneider_electrically_2013},
to fast and low-power switching \cite{Amo2010,dreismann_sub-femtojoule_2016}.
With a stronger polariton nonlinearity, polariton blockade \cite{MunozMatutano2019,Delteil2019}
and all-optical integrated quantum gates \cite{Ghosh2020} may also
be possible.

While nonlinearity plays a pivotal role in polaritonic phenomena,
it has been found to be relatively weak in commonly studied systems,
and its origin, controversial. The full solution of the polariton
interaction is a formidable quantum mechanics challenge, as we need
to solve a six-body problem involves two photons, two electrons and
two holes. Instead, most previous studies use the Born approximation,
or a mean-field approach \cite{NoteBornApproximation}. The polariton
interaction strength $g_{PP}$ is considered to be directly determined
by that of the exciton's, $g_{XX}$, as: $g_{PP}=X_{LP}^{4}g_{XX}$,
for $X_{LP}^{2}$ the Hopfield coefficient, corresponding to the exciton
fraction in the lower polariton (LP) mode. Treating the exciton scattering
in the Born approximation leads to the widely used result \citep{Ciuti1998,Tassone1999,Glazov2009,Levinsen2019}:
\begin{equation}
g_{PP}^{(0)}=X_{LP}^{4}g_{XX}^{(0)}\simeq X_{LP}^{4}\left(6.06E_{X}a_{X}^{2}\right),\label{eq:gPP0}
\end{equation}
where $E_{X}\equiv\hbar^{2}/(2m_{r}a_{X}^{2})$ and $a_{X}$ are the
binding energy and Bohr radius of excitons with a total mass $m_{X}=m_{e}+m_{h}$
and a reduced mass $m_{r}=m_{e}m_{h}/m_{X}$. However, there is a
fundamental conceptual inconsistency. Born approximation, indicated
here by the superscript ``$0$'', is often used in three dimensions.
But it is known to fail in low dimensions even at the \emph{qualitative}
level, due to strong quantum fluctuations \cite{Popov1972,Mora2009,He2015,Salasnich2016}

In this work, taking into account Gaussian quantum fluctuations in
a many-body approach \cite{Popov1972,Salasnich2016}, we obtain an
analytical expression for exciton and polariton interactions in two
dimensions (2D) beyond the Born approximation. We show that, while
the two-body exciton interaction strength $g_{XX}$ vanishes in 2D
due to quantum fluctuations \cite{Popov1972,Mora2009}, strong coupling
with photon introduces a new energy scale and leads to a finite two-body
polariton interaction strength $g_{PP}$ of the form: 
\begin{equation}
g_{PP}=X_{LP}^{4}\left(\frac{4\pi\hbar^{2}}{m_{X}}\right)\ln^{-1}\left[\frac{2}{e^{2\gamma}}\frac{\hbar^{2}}{m_{X}a_{s}^{2}\left|E_{LP}\right|}\right],\label{eq:gPPBog}
\end{equation}
where $\gamma\simeq0.577$ is Euler's constant, $a_{s}$ is the exciton-exciton
$s$-wave scattering length, and $E_{LP}=\delta/2-\sqrt{\delta^{2}/4+\Omega^{2}}$
is the lower polariton energy for photon-exciton detuning $\delta$
and coupling strength $\Omega$ \citep{Deng2010}. We furthermore
derive the exact universal Tan relations \cite{Tan2008a,Tan2008b,Braaten2008}
for 2D polaritons and determine Tan contact coefficient $\mathcal{I}$,
which underlies a $q^{-4}$ tail in the excitonic momentum distribution
$n_{X}(q\rightarrow\infty)\sim\mathcal{I}/q^{4}$.

Our results reveal that, contrary to the predication Eq.~(\ref{eq:gPP0}),
the polariton interaction strength may be greatly enhanced at negative
photon-exciton detuning when the exciton scattering length is large,
with correspondingly an even more dramatic increase in the Tan contact
coefficient $\mathcal{I}$. These predictions could be experimentally
checked in quantum wells \cite{Estrecho2019,Ferrier2011} or van der
Waals monolayers \cite{Tan2019arXiv,Emmanuele2019arXiv} placed in
microcavities. The unusual detuning dependence of the polariton-polariton
interaction strength could provide a way to measure the hitherto unknown
2D exciton-exciton scattering lengths \cite{SM} and to achieve strong
polariton nonlinearities in systems with large scattering lengths.

Equation (2) is applicable when excitons can be well regarded as point-like,
structureless bosons as in the standard exciton-polariton model, a
picture generally adopted by the polariton community \cite{Deng2010,Carusotto2013}.
We use the zero-temperature Bogoliubov theory \cite{Popov1972,Salasnich2016}
as a minimal description of those bosons, by assuming all photons
and excitons are coherently condensed into the zero-momentum state.
The strong Gaussian fluctuations are then well-characterized by Bogoliubov
quasiparticles out of the condensate. In this description, the interaction
energy is simply the chemical potential $\mu$ measured with respect
to $E_{LP}$, i.e., $E_{\textrm{int}}=\mu-E_{LP}=g_{PP}n$, which
is in turn proportional to the density $n$ in the dilute limit. We
will take advantage of this relation to calculate the two-body polariton-polariton
interaction strength $g_{PP}$, although we solve a many-body problem.

\textit{Model Hamiltonian}. The 2D electron-hole-photon system in
microcavities can be described by the following model Hamiltonian
$\mathscr{H}=\mathscr{H}_{0}+\mathscr{H}_{\textrm{LM}}+\mathscr{H}_{\textrm{int}}$
as 
\begin{eqnarray}
\mathscr{H}_{0} & = & \sum_{\mathbf{q}}\left[\left(\frac{\hbar^{2}\mathbf{q}^{2}}{2m_{\textrm{ph}}}+\delta-\mu\right)\phi_{\mathbf{q}}^{\dagger}\phi_{\mathbf{q}}+\xi_{\mathbf{q}}X_{\mathbf{q}}^{\dagger}X_{\mathbf{q}}\right],\\
\mathscr{H}_{\textrm{LM}} & = & \frac{\Omega}{\sqrt{\mathcal{S}}}\sum_{\mathbf{q}}\left[\phi_{\mathbf{q}}^{\dagger}X_{\mathbf{q}}+X_{\mathbf{q}}^{\dagger}\phi_{\mathbf{q}}\right],\\
\mathscr{H}_{\textrm{int}} & = & \frac{g}{2\mathcal{S}}\sum_{\mathbf{q}\mathbf{q}'\mathbf{k}}X_{\frac{\mathbf{k}}{2}+\mathbf{q}}^{\dagger}X_{\frac{\mathbf{k}}{2}-\mathbf{q}}^{\dagger}X_{\frac{\mathbf{k}}{2}-\mathbf{q}'}X_{\frac{\mathbf{k}}{2}+\mathbf{q}'}.\label{eq: HamiCoul}
\end{eqnarray}
Here, $\xi_{\mathbf{q}}\equiv\hbar^{2}\mathbf{q}^{2}/(2m_{X})-\mu$
is the excitonic dispersion relation with the chemical potential $\mu$
($<\delta$), $\mathcal{S}$ is the area of the system and hereafter
is taken to be unity, $\phi_{\mathbf{q}}$ and $X_{\mathbf{q}}$ are
the annihilation field operators for photons and excitons, respectively.
The mass of cavity photons $m_{\textrm{ph}}$ is typically several
orders smaller than the exciton mass $m_{X}$. In the interaction
Hamiltonian $\mathscr{H}_{\textrm{int}}$, $g$ is a \emph{bare} exciton
interaction strength, which is to be replaced by the exciton-exciton
$s$-wave scattering length $a_{s}$ according to \cite{Popov1972,Salasnich2016},
\begin{equation}
\frac{1}{g}+\sum_{\mathbf{q}}\left[\frac{\hbar^{2}\mathbf{q}^{2}}{m_{X}}+\varepsilon_{c}\right]^{-1}=\frac{m_{X}}{4\pi\hbar^{2}}\ln\left[\frac{4}{e^{2\gamma}}\frac{\hbar^{2}}{m_{X}a_{s}^{2}\varepsilon_{c}}\right].
\end{equation}
Here $\varepsilon_{c}>0$ is an arbitrary energy used to regularize
the \emph{infrared} divergence, which is unavoidable in 2D \cite{Popov1972,Salasnich2016}.

In the absence of the photon field, the model Hamiltonian describes
a weakly interacting 2D Bose gas and has been solved by Popov \cite{Popov1972},
based on whose work the density equation of state within the Bogoliubov
approximation was obtained as \cite{Mora2009,Salasnich2016,SM}: 
\begin{equation}
n\left(\mu_{B}\right)=\frac{m_{X}\mu_{B}}{4\pi\hbar^{2}}\ln\left[\frac{4}{e^{2\gamma+1}}\frac{\hbar^{2}}{m_{X}\mu_{B}a_{s}^{2}}\right].\label{eq:densityEoSExciton}
\end{equation}
This implies that the effective interaction strength $g_{XX}=\mu_{B}/n$
depends logarithmically on the chemical potential $\mu_{B}$ or the
density $n$, and consequently vanishes identically in the dilute
limit (i.e., $n\rightarrow0$). When we add the photon field, coherent
superposition of photons and excitons gives rise to two polariton
branches in the energy spectrum \citep{Deng2010}. Focusing on LP
only, the creation field operator can be written as $P^{\dagger}\simeq\sqrt{1-X_{LP}^{2}}\phi^{\dagger}+X_{LP}X^{\dagger}$
\citep{Deng2010}. As there is no interaction between photons, the
interaction between polaritons should come from the excitonic part.
By rewriting the interaction Hamiltonian $\mathscr{H}_{\textrm{int}}$
in terms of $P$ and $P^{\dagger}$, we then have the naive expression
$g_{PP}\simeq X_{LP}^{4}g_{XX}$, as we already see in Eq. (\ref{eq:gPP0})
within the Born approximation. Beyond the Born approximation, $g_{PP}$
therefore should vanish in the dilute limit, exactly in the same way
as the effective exciton interaction strength $g_{XX}$. This disagrees
with experimental findings \citep{Ferrier2011,Estrecho2019}. To solve
this apparent contradiction, we note that there could be virtual excitations
from the LP branch to the upper-polariton branch, by the residual
scattering terms (generated when we rewrite $\mathscr{H}_{\textrm{int}}$
in terms of $P$ and $P^{\dagger}$). These virtual excitations may
render the polariton-polariton interaction strength finite as we show
below using the Bologiubov theory.

\textit{Bogoliubov theory.} At zero temperature $T=0$, both photons
and excitons macroscopically condense into zero-momentum states with
wave-functions $\phi_{0}$ and $X_{0}$, respectively. To the leading
order, the mean-field thermodynamic potential takes the form, 
\begin{equation}
\varOmega_{0}\left(\mu\right)=\left(\delta-\mu\right)\phi_{0}^{2}+2\Omega\phi_{0}X_{0}-\mu X_{0}^{2}+\frac{g}{2}X_{0}^{4}.
\end{equation}
By minimizing $\varOmega_{0}(\mu)$, we obtain $gX_{0}^{2}=\mu+\Omega^{2}/(\delta-\mu)$
and $\varOmega_{0}(\mu)=-[\mu+\Omega^{2}/(\delta-\mu)]^{2}/(2g).$

To take into account crucial quantum fluctuations in 2D, we rewrite
the Hamiltonian in terms of $\delta\phi=\phi-\phi_{0}$ and $\delta X=X-X_{0}$
and keep only the bilinear terms at the Gaussian level. We then obtain
the inverse Green function $\mathscr{D}^{-1}(\mathbf{q},i\nu_{n})$
of the Bogoliubov quasiparticles \citep{Salasnich2016}, 
\[
\mathscr{-D}^{-1}=\left[\begin{array}{cccc}
-i\nu_{n}+A_{\mathbf{q}} & 0 & \Omega & 0\\
0 & i\nu_{n}+A_{\mathbf{q}} & 0 & \Omega\\
\Omega & 0 & -i\nu_{n}+B_{\mathbf{q}} & C\\
0 & \Omega & C & i\nu_{n}+B_{\mathbf{q}}
\end{array}\right],
\]
where $\nu_{n}\equiv2\pi nk_{B}T$ ($n\in\mathbb{Z}$) are bosonic
Matsubara frequencies and we have introduced the notations, 
\begin{eqnarray}
A_{\mathbf{q}} & \equiv & \frac{\hbar^{2}\mathbf{q}^{2}}{2m_{\textrm{ph}}}+\delta-\mu,\\
B_{\mathbf{q}} & \equiv & \frac{\hbar^{2}\mathbf{q}^{2}}{2m_{X}}-\mu+2gX_{0}^{2}=\frac{\hbar^{2}\mathbf{q}^{2}}{2m_{X}}+\mu+\frac{2\Omega^{2}}{\delta-\mu},\\
C & \equiv & gX_{0}^{2}=\mu+\frac{\Omega^{2}}{\delta-\mu}.
\end{eqnarray}
As the lowest attainable chemical potential is $E_{LP}$, i.e., $\mu=E_{LP}+\mu_{B}$
where $\mu_{B}>0$, in the dilute limit we find $C\simeq[1+\Omega^{2}/(\delta-E_{LP})^{2}]\mu_{B}=X_{LP}^{-2}\mu_{B}>0.$
By solving $\det\left[\mathscr{D}^{-1}(\mathbf{q},i\nu_{n}\rightarrow E)\right]=0$,
we obtain the quasiparticle energy spectrum, 
\[
E_{\mathbf{q}\pm}^{2}=\mathcal{K}_{\mathbf{q+}}+\Omega^{2}\pm\sqrt{\mathcal{K}_{\mathbf{q}-}^{2}+\left[\left(A_{\mathbf{q}}+B_{\mathbf{q}}\right)^{2}-C^{2}\right]\Omega^{2}},
\]
where we have defined $\mathcal{K}_{\mathbf{q\pm}}\equiv[A_{\mathbf{q}}^{2}\pm B_{\mathbf{q}}^{2}\mp C^{2}]/2$.

At the Gaussian level for quantum fluctuations, quasiparticles are
approximately treated as \emph{non-interacting} particles. Thus, their
contribution to the thermodynamic potential can be written down straightforwardly
\citep{AGD1963}, 
\begin{equation}
\delta\varOmega_{g}=\frac{k_{B}T}{2}\sum_{\mathbf{q},i\nu_{n}}\ln\det\left[\mathscr{-D}^{-1}(\mathbf{q},i\nu_{n})\right]e^{i\nu_{n}0^{+}},\label{eq:dOmega}
\end{equation}
where the convergence factor $e^{i\nu_{n}0^{+}}$ is used to regularize
the divergence at $\nu_{n}\rightarrow\pm\infty$. As we discuss in
detail in Supplemental Material \citep{SM}, the summation over the
bosonic Matsubara frequencies can be explicitly performed and at zero
temperature we find $\delta\varOmega_{g}^{(T=0)}=\sum_{\mathbf{q}}[E_{\mathbf{q}+}+E_{\mathbf{q}-}-A_{\mathbf{q}}-B_{\mathbf{q}}]/2$,
which formally diverges. However, this ultraviolet divergence can
be exactly cancelled by the same divergence in the mean-field thermodynamic
potential $\varOmega_{0}$. By putting these two contributions together,
i.e., $\varOmega=\varOmega_{0}+\delta\varOmega_{g}^{(T=0)}$, we arrive
at \cite{SM}
\begin{eqnarray}
\varOmega & = & -\frac{m_{X}C^{2}}{8\pi\hbar^{2}}\ln\left[\frac{4}{e^{2\gamma}}\frac{\hbar^{2}}{m_{X}a_{s}^{2}\varepsilon_{c}}\right]+\frac{1}{2}\sum_{\mathbf{q}}\left[E_{\mathbf{q}+}+E_{\mathbf{q}-}\right.\nonumber \\
 &  & \left.-A_{\mathbf{q}}-B_{\mathbf{q}}+\frac{C^{2}}{\hbar^{2}q^{2}/m_{X}+\varepsilon_{c}}\right].\label{eq:OmegaBog}
\end{eqnarray}

At nonzero light-matter coupling, interestingly, the integration over
the momentum in the above can be worked out analytically in the infinite
mass ratio limit $m_{X}/m_{\textrm{ph}}\rightarrow\infty$. We find
that \citep{SM}, 
\begin{equation}
\varOmega=-\frac{m_{X}}{8\pi\hbar^{2}}\left[\mu+\frac{\Omega^{2}}{\delta-\mu}\right]^{2}\ln\left[\frac{2}{e^{2\gamma}}\frac{\hbar^{2}\left(\delta-\mu\right)}{m_{X}a_{s}^{2}\Omega^{2}}\right].\label{eq:OmegaBog1}
\end{equation}
By keeping the leading term in powers of $\mu_{B}=\mu-E_{LP}$ and
taking derivative of $\varOmega$ with respect to $\mu_{B}$, i.e.,
$n=-\partial\varOmega/\partial\mu_{B}$, we obtain 
\begin{equation}
n=\frac{\mu_{B}}{X_{LP}^{4}}\left(\frac{m_{X}}{4\pi\hbar^{2}}\right)\ln\left[\frac{2}{e^{2\gamma}}\frac{\hbar^{2}}{m_{X}a_{s}^{2}\left|E_{LP}\right|}\right]
\end{equation}
and hence the polariton-polariton interaction strength in Eq. (\ref{eq:gPPBog}).
By comparing the above density equation with Eq. (\ref{eq:densityEoSExciton}),
we see that the small chemical potential $\mu_{B}$ in the logarithm
is now replaced with a characteristic finite LP energy, due to the
virtual scatterings between the two polariton branches. As a result,
the polariton-polariton interaction strength in Eq. (\ref{eq:gPPBog})
becomes finite in the dilute limit. This observation is the first
main result of our work. It is also applicable to the case of $N$
quantum wells, where the polariton interaction is reduced by a factor
of $N$ \citep{SM}.

\begin{figure}[t]
\begin{centering}
\includegraphics[width=0.45\textwidth]{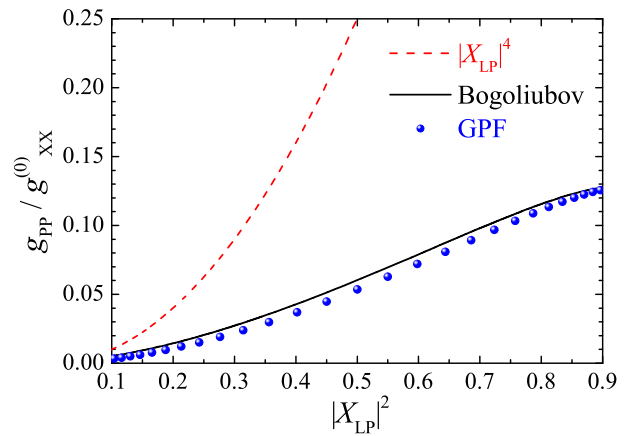} 
\par\end{centering}
\centering{}\caption{\label{fig1_BogGPF} Comparison of the polariton-polariton interaction
strength $g_{PP}$ from the Bogoliubov theory (solid line), the fermionic
toy model based on GPF theory (blue circles) and the Born approximation
(red dashed line), showing good agreement between the Bogoliubov theory
and GPF theory predictions. Here $g_{PP}$ is measured in units of
$g_{XX}^{(0)}=4\pi\hbar^{2}/M$, as a function of $X_{LP}^{2}$ at
$\Omega=0.1E_{X}$. }
\end{figure}

\textit{Validity of our results. } Eq. (\ref{eq:gPPBog}) is an exact
\emph{two-body} result, valid as long as the \emph{bosonic} model
holds. This implies that we need $\Omega<\Omega_{c}\ll E_{X}$ and
the density $n<n_{c}\sim0.01a_{X}^{-2}$, so that the internal fermionic
degrees of freedom of excitons are frozen and do not lead to observable
effects. To estimate $\Omega_{c}$, we compare our results with a
\emph{fermionic} toy model without such a restriction, where the Coulomb
interaction is approximately replaced by a contact interaction and
electrons and holes are assumed to have the same mass $m_{e}=m_{h}=M$.
It can be reliably solved by using a fermionic Gaussian pair fluctuation
(GPF) theory, which in the dilute limit recovers the bosonic Bogoliubov
theory \citep{Hu2006,Hu2020arXiv}. Within the toy model, the exciton
$s$-wave scattering length $a_{s}\simeq1.12e^{-\gamma}a_{X}$ is
known \citep{He2015}. Therefore, we can compare the predictions from
both the bosonic and fermionic models under the same condition. As
shown in Fig. \ref{fig1_BogGPF}, we find a good agreement at $\Omega=0.1E_{X}$,
indicating $\Omega_{c}\sim0.1E_{X}$.

Experimentally, the polariton interaction has been reported for MoSe$_{2}$
monolayers at $\Omega=5.0$ meV \citep{Tan2019arXiv} or $\Omega=17.2$
meV \citep{Emmanuele2019arXiv} near zero detuning. These light-matter
couplings are much smaller than the exciton binding energy $E_{X}\sim500$
meV \citep{Wang2018}. Using Eq. (\ref{eq:gPPBog}) and $m_{e}\simeq m_{h}\simeq0.5m_{0}$
for MoSe$_{2}$ \citep{Wang2018}, we obtain $g_{PP}\sim0.1\mu\textrm{eV}\cdot\mu\textrm{m}^{2}$,
which is consistent with the experimental data $g_{PP}=0.01-1.0\mu\textrm{eV}\cdot\mu\textrm{m}^{2}$
\citep{Tan2019arXiv,Emmanuele2019arXiv}.

\begin{figure}[t]
\begin{centering}
\includegraphics[width=0.45\textwidth]{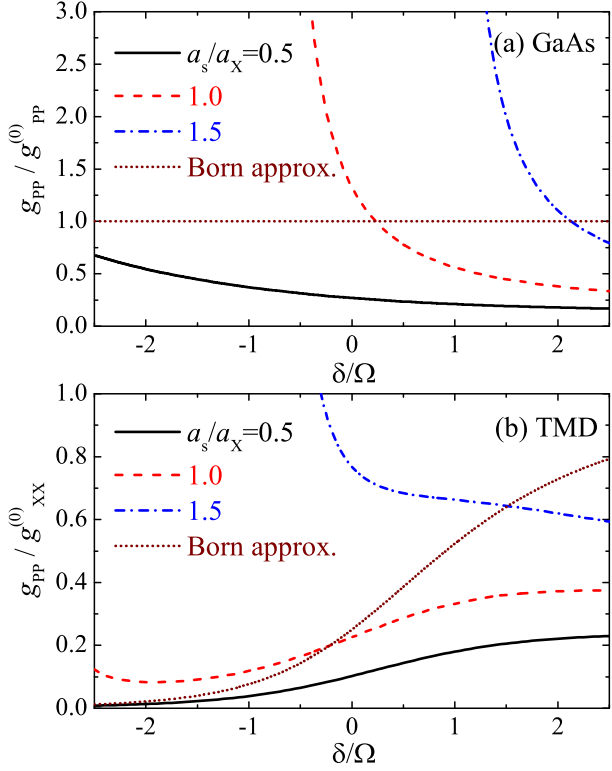} 
\par\end{centering}
\centering{}\caption{\label{fig2_gpp} The detuning dependence of $g_{PP}$, relative to
$g_{PP}^{(0)}=X_{LP}^{4}g_{XX}^{(0)}$ for GaAs quantum well (a) and
relative to $g_{XX}^{(0)}$ for TMD monolayer (b), at different scattering
lengths $a_{s}/a_{X}=0.5$ (solid line), $1.0$ (dashed line) and
$1.5$ (dot-dashed line), and at $\Omega=0.1E_{X}$. The dotted line
is the Born approximation result, $g_{PP}^{(0)}/g_{XX}^{(0)}=X_{LP}^{4}$.}
\end{figure}

\textit{Anomalous interaction enhancement}. The inverse logarithmic
dependence of the polariton-polariton interaction strength $g_{PP}$
on the LP energy $E_{LP}$ shown in Eq. (\ref{eq:gPPBog}) is nontrivial.
As $\left|E_{LP}\right|$ can be enlarged by tuning the photon detuning
even at $\Omega<\Omega_{c}$, we find the second main result of our
work that the polariton interaction could be anomalously enhanced
at a large \emph{red} detuning. To see this, for the Coulomb interaction
let us recast the expression of $g_{PP}$ into the form, 
\begin{equation}
\frac{g_{PP}}{g_{XX}^{(0)}}=\frac{4\pi}{3.03}\frac{m_{r}}{m_{X}}\frac{X_{LP}^{4}}{\left[-\ln\left(\left|E_{LP}\right|/E_{X}\right)+\mathcal{C}_{PP}\right]},
\end{equation}
where $\mathcal{C}_{PP}\equiv2\ln(a_{s}/a_{X})+\ln(m_{r}/m_{X})+2\gamma-2\ln2$.
Clearly, a resonance appears at $\left|E_{LP}\right|=\mathcal{C}_{PP}E_{X}$,
when the photon field is significantly occupied and the scattering
between excitons is then drastically altered. Our perturbative Bogoliubov
theory breaks down at resonance. However, away from the resonance
the qualitative anomalous enhancement seems to be physical.

In Fig. \ref{fig2_gpp}, we report the polariton-polariton interaction
strengths for GaAs quantum well (a) and TMD monolayer (b) in microcavities,
with masses $m_{e}\simeq0.067m_{0}$ and $m_{h}\simeq0.45m_{0}$ (GaAs)
\citep{Deng2010} and $m_{e}\simeq m_{h}\simeq0.5m_{0}$ (TMD) \citep{Wang2018},
respectively. As the exciton-exciton $s$-wave scattering length $a_{s}$
remains elusive for the Coulomb interaction in 2D \cite{SM} and might
be tunable \cite{SM,Cudazzo2011}, we consider three likely choices,
as inspired by the result in three dimensions (i.e., $a_{s}\sim a_{X}$)
\citep{Golomedov2017}. In comparison to the Born approximation result
$g_{PP}^{(0)}$, as shown in Fig. \ref{fig2_gpp}(a), we find the
ratio $g_{PP}/g_{PP}^{(0)}$ decreases monotonically with increasing
photon detuning. In contrast, measured in units of $g_{XX}^{(0)}$
as plotted in Fig. \ref{fig2_gpp}(b), the anomalous enhancement becomes
less apparent, except at large $a_{s}\sim1.5a_{X}$ where the rise
at red detuning is always significant. This sensitive dependence of
the polariton interaction on $a_{s}$ provides a unique way to measure
the long-sought exciton-exciton scattering length in 2D semiconductor
materials \cite{SM}.

\begin{figure}[t]
\begin{centering}
\includegraphics[width=0.45\textwidth]{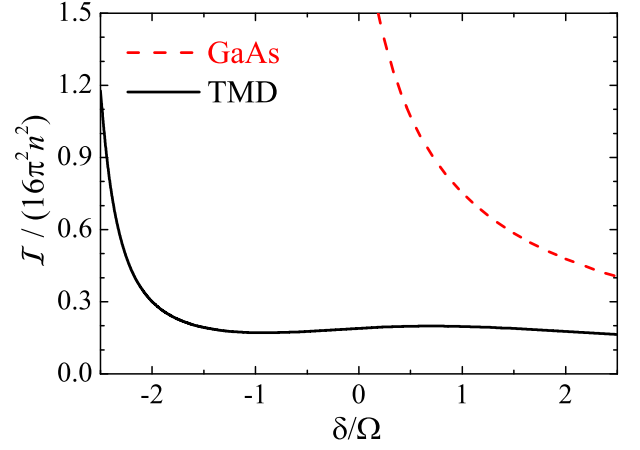} 
\par\end{centering}
\centering{}\caption{\label{fig3_contact} Tan's contact coefficient (in units of $16\pi^{2}n^{2}$)
as a function of the detuning at $\Omega=0.1E_{X}$, for GaAs quantum
well (dashed line) and TMD monolayer (solid line). The scattering
length $a_{s}$ is set to be $a_{X}$.}
\end{figure}

\textit{Tan relations}. We now consider the universal relations which
govern the short-range, large-momentum and high-energy behaviors of
a quantum many-body system \cite{Tan2008a,Tan2008b,Braaten2008}.
In these exact relations, the central role is played by Tan's contact
coefficient $\mathcal{I}=(m_{X}^{2}g^{2}/\hbar^{4})\int d\mathbf{r}\left\langle X^{\dagger}(\mathbf{r})X^{\dagger}(\mathbf{r})X(\mathbf{r})X(\mathbf{r})\right\rangle $.
As discussed in detail in Supplemental Material, we derive the adiabatic
and energy relations \cite{SM}, 
\begin{eqnarray}
\left[\frac{\partial\varOmega}{\partial\ln a_{s}}\right]_{\mu,S} & = & \frac{\hbar^{2}}{4\pi m_{X}}\mathcal{I},\label{eq:AdibaticRelation}\\
\mathscr{T}_{X}+\mathscr{E}_{\textrm{int}} & = & \sum_{\mathbf{q}}\frac{\hbar^{2}q^{2}}{2m_{X}}\tilde{n}_{X}\left(q\right)+\frac{\left(\ln2-\gamma\right)\hbar^{2}\mathcal{I}}{4\pi m_{X}},\label{eq:EnergyRelation}
\end{eqnarray}
where $\tilde{n}_{X}(q)\equiv n_{X}(q)-\mathcal{I}/[q^{2}(q^{2}+a_{s}^{-2})]$,
$\mathscr{T}_{X}$ and $\mathscr{E}_{\textrm{int}}$ are the excitonic
kinetic energy and interaction energy, respectively. By applying the
adiabatic relation to Eq. (\ref{eq:OmegaBog1}), we obtain within
the Bogoliubov approximation, 
\begin{equation}
\mathcal{I}=\frac{m_{X}^{2}C^{2}}{\hbar^{4}}\simeq\frac{\left(16\pi^{2}n^{2}\right)X_{LP}^{4}}{\ln^{2}\left[e^{2\gamma}m_{X}a_{s}^{2}\left|E_{LP}\right|/\left(2\hbar^{2}\right)\right]}.
\end{equation}
Figure \ref{fig3_contact} presents the detuning dependence of the
contact coefficient for GaAs quantum well (dashed line) and TMD monolayer
(solid line) at $a_{s}=a_{X}$. In accordance with the anomalous enhancement
in the polariton interaction, we also observe a dramatic increase
in the contact coefficient at red detuning, which can be measured
from the universal $q^{-4}$ tail in the excitonic momentum distribution
$n_{X}(q)$.

\textit{Conclusions}. We have derived an analytic expression for interactions
of two-dimensional exciton-polaritons. Compared to the previous constant
two-body interaction strength derived within the Born approximation,
our result shows a logarithmic dependence on both the exciton $s$-wave
scattering length and shift of the polariton energy from the bare
exciton energy. Such a dependence leads to a counter intuitive, large
enhancement in the polariton-polariton interaction strength and Tan's
contact coefficient at red photon-exciton detunings when the scattering
length is greater than the exciton Bohr radius. Therefore our result
suggests a way to measure the 2D exciton scattering length and reveals
the possibility of achieving a strongly nonlinear polariton gas in
materials with a large exciton scattering length. 
\begin{acknowledgments}
This research was supported by the Australian Research Council's (ARC)
Discovery Program, Grant No. DP170104008 (H.H.) and Grant No. DP180102018
(X.-J.L), and by the Army Research Office under Awards W911NF-17-1-0312
(H.D.). 
\end{acknowledgments}

\appendix

\begin{widetext}

\section{Quantum fluctuation thermodynamic potential}

At the Gaussian level for quantum fluctuations, Bogoliubov quasiparticles
are treated as non-interacting and described by the Green function,
\begin{equation}
\mathscr{-D}^{-1}\left(\mathbf{q},i\nu_{n}\right)=\left[\begin{array}{cccc}
-i\nu_{n}+A_{\mathbf{q}} & 0 & \Omega & 0\\
0 & i\nu_{n}+A_{\mathbf{q}} & 0 & \Omega\\
\Omega & 0 & -i\nu_{n}+B_{\mathbf{q}} & C\\
0 & \Omega & C & i\nu_{n}+B_{\mathbf{q}}
\end{array}\right],
\end{equation}
where $\nu_{n}\equiv2\pi nk_{B}T$ ($n\in\mathbb{Z}$) are bosonic
Matsubara frequencies, and 
\begin{eqnarray}
A_{\mathbf{q}} & \equiv & \frac{\hbar^{2}\mathbf{q}^{2}}{2m_{\textrm{ph}}}+\delta-\mu,\\
B_{\mathbf{q}} & \equiv & \frac{\hbar^{2}\mathbf{q}^{2}}{2m_{X}}-\mu+2gX_{0}^{2}=\frac{\hbar^{2}\mathbf{q}^{2}}{2m_{X}}+\mu+\frac{2\Omega^{2}}{\delta-\mu},\\
C & \equiv & gX_{0}^{2}=\mu+\frac{\Omega^{2}}{\delta-\mu}.
\end{eqnarray}
We note that the chemical potential satisfies 
\begin{equation}
\delta>\mu>E_{LP}=\frac{\delta}{2}-\sqrt{\frac{\delta^{2}}{4}+\Omega^{2}}.
\end{equation}
As a result, we have $A_{\mathbf{q}}>0$, $B_{\mathbf{q}}>0$ and
$C>0$. In particular, by writing $\mu=E_{LP}+\mu_{B}$ with $\mu_{B}>0$,
we find 
\begin{equation}
C\simeq\left[1+\frac{\Omega^{2}}{\left(\delta-E_{LP}\right)^{2}}\right]\mu_{B}=\frac{\mu_{B}}{X_{LP}^{2}}
\end{equation}
in the dilute zero-density limit (i.e., $\mu_{B}\rightarrow0$). The
poles of the Green function give the energy spectrum of Bogoliubov
quasiparticles. We therefore solve the eigenvalue equation,
\begin{eqnarray}
\det\left[\mathscr{D}^{-1}\left(\mathbf{q},i\nu_{n}\rightarrow E\right)\right] & =E^{4}-\left(A_{\mathbf{q}}^{2}+B_{\mathbf{q}}^{2}-C^{2}+2\Omega^{2}\right)E^{2}+\left(A_{\mathbf{q}}^{2}B_{\mathbf{q}}^{2}-A_{\mathbf{q}}^{2}C^{2}-2A_{\mathbf{q}}B_{\mathbf{q}}\Omega^{2}+\Omega^{4}\right)= & 0,
\end{eqnarray}
and find the quasiparticle energy spectrum,

\begin{equation}
E_{\mathbf{q}\pm}^{2}=\left(\frac{A_{\mathbf{q}}^{2}+B_{\mathbf{q}}^{2}-C^{2}}{2}+\Omega^{2}\right)\pm\sqrt{\left(\frac{A_{\mathbf{q}}^{2}-B_{\mathbf{q}}^{2}+C^{2}}{2}\right)^{2}+\left[\left(A_{\mathbf{q}}+B_{\mathbf{q}}\right)^{2}-C^{2}\right]\Omega^{2}}.
\end{equation}
It is easy to check that at zero momentum $\mathbf{q}=0$, the lower
spectrum $E_{\mathbf{q}-}=0$. This is anticipated, as the quasiparticle
spectrum must have a gapless Goldstone model, as a result of the $U(1)$
symmetry breaking.

For \emph{non-interacting} bosons, their thermodynamic potential takes
the form \cite{AGD1963},

\begin{equation}
\delta\varOmega_{g}=\frac{k_{B}T}{2}\sum_{\mathbf{q},i\nu_{n}}\ln\det\left[\mathscr{-D}^{-1}\left(\mathbf{q},i\nu_{n}\right)\right]e^{i\nu_{n}0^{+}}=\frac{k_{B}T}{2}\sum_{\mathbf{q},i\nu_{n}}\ln\left[\left(\nu_{n}^{2}+E_{\mathbf{q}+}^{2}\right)\left(\nu_{n}^{2}+E_{\mathbf{q}-}^{2}\right)\right]e^{i\nu_{n}0^{+}}.\label{eq:dOmega-1}
\end{equation}
Here, it is necessary to add the convergence factor $e^{i\nu_{n}0^{+}}$
to regularize the ultraviolet divergence at $\nu_{n}\rightarrow\pm\infty$.
This is required even for the simplest case of single-component non-interacting
bosons with dispersion relation $\xi_{\mathbf{q}}=\hbar^{2}\mathbf{q}^{2}/(2M)-\mu>0$,
where the thermodynamic potential is known as,
\begin{equation}
\varOmega_{B}=\frac{k_{B}T}{2}\sum_{\mathbf{q},i\nu_{n}}\ln\left[\nu_{n}^{2}+\xi_{\mathbf{q}}^{2}\right]e^{i\nu_{n}0^{+}}=k_{B}T\sum_{\mathbf{q},i\nu_{n}}\ln\left[i\nu_{n}-\xi_{\mathbf{q}}\right]e^{i\nu_{n}0^{+}}=\frac{1}{\exp\left[\xi_{\mathbf{q}}/\left(k_{B}T\right)\right]-1}\overset{\textrm{if \ensuremath{T=0}}}{=}0.
\end{equation}
Let us now subtract this zero contribution, i.e., 
\begin{equation}
\frac{k_{B}T}{2}\sum_{\mathbf{q},i\nu_{n}}\ln\left[\left(\nu_{n}^{2}+A_{\mathbf{q}}^{2}\right)\left(\nu_{n}^{2}+B_{\mathbf{q}}^{2}\right)\right]e^{i\nu_{n}0^{+}}=0
\end{equation}
from the thermodynamic potential $\delta\varOmega_{g}^{(T=0)}$. We
obtain,
\begin{equation}
\delta\varOmega_{g}^{(T=0)}=\frac{k_{B}T}{2}\sum_{\mathbf{q},i\nu_{n}}\ln\left[\frac{\left(\nu_{n}^{2}+E_{\mathbf{q}+}^{2}\right)\left(\nu_{n}^{2}+E_{\mathbf{q}-}^{2}\right)}{\left(\nu_{n}^{2}+A_{\mathbf{q}}^{2}\right)\left(\nu_{n}^{2}+B_{\mathbf{q}}^{2}\right)}\right],
\end{equation}
where the convergence factor has been removed, as the integrand now
vanishes in the limit $\nu_{n}\rightarrow\pm\infty$ and the integral
converges. At \emph{zero} temperature, by using the identity (i.e.,
$\nu_{n}\rightarrow\omega$)
\begin{equation}
k_{B}T\sum_{i\nu_{n}}\ln\left[\frac{\nu_{n}^{2}+E^{2}}{\nu_{n}^{2}+\xi^{2}}\right]=\frac{1}{2\pi}\intop_{-\infty}^{+\infty}d\omega\left[\frac{\omega^{2}+E^{2}}{\omega^{2}+\xi^{2}}\right]=E-\xi,
\end{equation}
we obtain,
\begin{equation}
\delta\varOmega_{g}^{(T=0)}=\frac{1}{2}\sum_{\mathbf{q}}\left[E_{\mathbf{q}+}+E_{\mathbf{q}-}-A_{\mathbf{q}}-B_{\mathbf{q}}\right].
\end{equation}

It is worth noting that the integrand in $\delta\varOmega_{g}^{(T=0)}$
is formally divergent. To see this, let us simply consider a zero
light-matter coupling $\Omega=0$, so the photon field is decoupled
from the exciton field. In this case, we find that $C=\mu>0$, $B_{\mathbf{q}}=\hbar^{2}\mathbf{q}^{2}/(2m_{X})+\mu$,
and 
\begin{eqnarray}
E_{\mathbf{q}+} & = & A_{\mathbf{q}},\\
E_{\mathbf{q}-} & = & \sqrt{\frac{\hbar^{2}\mathbf{q}^{2}}{2m_{X}}\left(\frac{\hbar^{2}\mathbf{q}^{2}}{2m_{X}}+2\mu\right)}.
\end{eqnarray}
Therefore, at large momentum the integrand will be 
\begin{equation}
E_{\mathbf{q}-}-B_{\mathbf{q}}=\sqrt{\frac{\hbar^{2}\mathbf{q}^{2}}{2m_{X}}\left(\frac{\hbar^{2}\mathbf{q}^{2}}{2m_{X}}+2\mu\right)}-\left(\frac{\hbar^{2}\mathbf{q}^{2}}{2m_{X}}+\mu\right)\simeq-\frac{1}{2}\frac{\mu^{2}}{\hbar^{2}\mathbf{q}^{2}/\left(2m_{X}\right)+\mu}.
\end{equation}
It is then easy to check the integral of $\delta\varOmega_{g}^{(T=0)}$
is logarithmically divergent. This divergence is actually anticipated,
as the mean-field Gross-Pitaevskii thermodynamic potential 
\begin{equation}
\varOmega_{0}=-\frac{\left[\mu+\Omega^{2}/\left(\delta-\mu\right)\right]^{2}}{2g}
\end{equation}
is equally logarithmically divergent. These two divergences will be
exactly cancelled once we add the two thermodynamic potentials together,
i.e., $\varOmega=\varOmega_{0}+\delta\varOmega_{g}^{(T=0)}$. By expressing
the \emph{bare} interaction strength $g$ in terms of the exciton-exciton
$s$-wave scattering length \cite{Salasnich2016}, i.e., 
\begin{equation}
\frac{1}{g}=\frac{m_{X}}{4\pi\hbar^{2}}\ln\left[\frac{4}{e^{2\gamma}}\frac{\hbar^{2}}{m_{X}a_{s}^{2}\varepsilon_{c}}\right]-\sum_{\mathbf{q}}\left[\frac{\hbar^{2}\mathbf{q}^{2}}{m_{X}}+\varepsilon_{c}\right]^{-1},\label{eq:gReg}
\end{equation}
we arrive at,

\begin{equation}
\varOmega=-\frac{m_{X}C^{2}}{8\pi\hbar^{2}}\ln\left[\frac{4}{e^{2\gamma}}\frac{\hbar^{2}}{m_{X}a_{s}^{2}\varepsilon_{c}}\right]+\frac{1}{2}\sum_{\mathbf{q}}\left[E_{\mathbf{q}+}+E_{\mathbf{q}-}-A_{\mathbf{q}}-B_{\mathbf{q}}+\frac{C^{2}}{\hbar^{2}q^{2}/m_{X}+\varepsilon_{c}}\right].\label{eq:OmegaBog-1}
\end{equation}

\subsection{Density equation of state of excitons}

To obtain an analytic expression for the thermodynamic potential $\varOmega$,
let us first check the case of excitons in the absence of the light-matter
coupling, $\Omega=0$. As mentioned earlier, we would have $C=\mu=\mu_{B}>0$.
By choosing a cut-off energy $\varepsilon_{c}=\mu_{B}$, we find that
the integral in $\varOmega$ is,
\begin{align}
\sum_{\mathbf{q}}\left[E_{\mathbf{q}+}+E_{\mathbf{q}-}-A_{\mathbf{q}}-B_{\mathbf{q}}+\frac{C^{2}}{\frac{\hbar^{2}q^{2}}{M_{X}}+\varepsilon_{c}}\right] & =\sum_{\mathbf{q}}\left[\sqrt{\frac{\hbar^{2}\mathbf{q}^{2}}{2m_{X}}\left(\frac{\hbar^{2}\mathbf{q}^{2}}{2m_{X}}+2\mu_{B}\right)}-\left(\frac{\hbar^{2}\mathbf{q}^{2}}{2m_{X}}+\mu_{B}\right)+\frac{\mu_{B}^{2}}{\frac{\hbar^{2}q^{2}}{m_{X}}+\mu_{B}}\right],\\
 & =\frac{m_{X}\mu_{B}^{2}}{8\pi\hbar^{2}}\intop_{0}^{\infty}dx\left[\sqrt{x\left(x+2\right)}-\left(x+1\right)+\frac{1}{2x+1}\right],\\
 & =\frac{m_{X}}{8\pi\hbar^{2}}\frac{\mu_{B}^{2}}{2},\label{eq:IntegralExcitons}
\end{align}
where in the second equation, we have introduced a dimensionless variable
$x\equiv\hbar^{2}\mathbf{q}^{2}/(2m_{X}\mu_{B})$. Therefore, we obtain
the thermodynamic potential \cite{Mora2009},
\begin{equation}
\varOmega=-\frac{m_{X}\mu_{B}^{2}}{8\pi\hbar^{2}}\ln\left[\frac{4}{e^{2\gamma}}\frac{\hbar^{2}}{m_{X}a_{s}^{2}\mu_{B}}\right]+\frac{1}{2}\left(\frac{m_{X}}{8\pi^{2}\hbar^{2}}\right)\mu_{B}^{2}=-\frac{m_{X}\mu_{B}^{2}}{8\pi\hbar^{2}}\ln\left[\frac{4}{e^{2\gamma+1/2}}\frac{\hbar^{2}}{m_{X}a_{s}^{2}\mu_{B}}\right].
\end{equation}
By taking the derivative with respect to the chemical potential $\mu_{B}$,
we obtain the density equation of state for an excitonic gas,
\begin{equation}
n=\frac{m_{X}\mu_{B}}{4\pi\hbar^{2}}\ln\left[\frac{4}{e^{2\gamma+1}}\frac{\hbar^{2}}{m_{X}a_{s}^{2}\mu_{B}}\right],
\end{equation}
which is already shown in the main text.

\subsection{Density equation of state of polaritons}

Let us now consider the thermodynamic potential in the presence of
the light-matter coupling $\Omega\neq0$ and in the limit of an infinitely
large mass ratio $m_{X}/m_{\textrm{ph}}\rightarrow\infty$. In this
limit, $A_{\mathbf{q}}$ is (infinitely) large for any nonzero momentum.
Therefore, we may approximate,
\begin{equation}
E_{\mathbf{q}\pm}^{2}\simeq\left(\frac{A_{\mathbf{q}}^{2}+B_{\mathbf{q}}^{2}-C^{2}}{2}+\Omega^{2}\right)\pm\left(\frac{A_{\mathbf{q}}^{2}-B_{\mathbf{q}}^{2}+C^{2}}{2}\right)\left\{ 1+\frac{2\Omega^{2}\left[\left(A_{\mathbf{q}}+B_{\mathbf{q}}\right)^{2}-C^{2}\right]}{\left(A_{\mathbf{q}}^{2}-B_{\mathbf{q}}^{2}+C^{2}\right)^{2}}\right\} .
\end{equation}
It is then easy to check that,
\begin{eqnarray}
E_{\mathbf{q}+}^{2} & \simeq & A_{\mathbf{q}}^{2}+\frac{2A_{\mathbf{q}}\left(A_{\mathbf{q}}+B_{\mathbf{q}}\right)}{A_{\mathbf{q}}^{2}-B_{\mathbf{q}}^{2}+C^{2}}\Omega^{2},\label{eq:EqpApprox}\\
E_{\mathbf{q}-}^{2} & \simeq & \left(B_{\mathbf{q}}^{2}-C^{2}\right)-\frac{2\left[\left(A_{\mathbf{q}}+B_{\mathbf{q}}\right)B_{\mathbf{q}}-C^{2}\right]}{A_{\mathbf{q}}^{2}-B_{\mathbf{q}}^{2}+C^{2}}\Omega^{2}.\label{eq:EqmApprox}
\end{eqnarray}
In the limit of $m_{X}/m_{\textrm{ph}}\gg1$, we may neglect the second
terms in $E_{\mathbf{q}\pm}^{2}$. In other words, the dispersion
relations of the photon field and exciton field are effectively decoupled,
although the excitonic dispersion is still strongly affected by the
light-matter coupling. Therefore, we find that,
\begin{equation}
\sum_{\mathbf{q}}\left[E_{\mathbf{q}+}+E_{\mathbf{q}-}-A_{\mathbf{q}}-B_{\mathbf{q}}+\frac{C^{2}}{\frac{\hbar^{2}q^{2}}{m_{X}}+\varepsilon_{c}}\right]\simeq\sum_{\mathbf{q}}\left[\sqrt{B_{\mathbf{q}}^{2}-C^{2}}-B_{\mathbf{q}}+\frac{C^{2}}{2B_{\mathbf{q}}}\right]+\sum_{\mathbf{q}}\left[\frac{C^{2}}{\frac{\hbar^{2}q^{2}}{m_{X}}+\varepsilon_{c}}-\frac{C^{2}}{2B_{q}}\right].
\end{equation}
The first integral can be casted into the form (i.e., $y=B_{\mathbf{q}}/C-1$),
\begin{equation}
\sum_{\mathbf{q}}\left[\sqrt{B_{\mathbf{q}}^{2}-C^{2}}-B_{\mathbf{q}}+\frac{C^{2}}{2B_{\mathbf{q}}}\right]=\frac{m_{X}C^{2}}{4\pi\hbar^{2}}\tilde{I}_{1},
\end{equation}
where the dimensionless integral $\tilde{I}_{1}$ is
\begin{equation}
\tilde{I}_{1}=2\intop_{\tilde{B}_{0}-1}^{\infty}dy\left[\sqrt{y\left(y+2\right)}-\left(y+1\right)+\frac{1}{2y+2}\right]
\end{equation}
and $\tilde{B}_{0}\equiv B_{\mathbf{q=0}}/C=1+\Omega^{2}/[(\delta-\mu)C]\geq1$.
Actually, with a nonzero light-matter coupling $\Omega\neq0$, $\tilde{B}_{0}\rightarrow+\infty$
in the dilute limit since $C\rightarrow0^{+}$. It is easy to check
that,
\begin{align}
\tilde{I}_{1} & =\left[\left(y+1\right)\sqrt{y\left(y+2\right)}-2\textrm{arcsinh}\sqrt{\frac{y}{2}}-y^{2}-2y+\ln\left(2y+2\right)\right]_{\tilde{B}_{0}-1}^{\infty},\\
 & =\left(\tilde{B}_{0}^{2}-\frac{1}{2}-\tilde{B}_{0}\sqrt{\tilde{B}_{0}^{2}-1}\right)+2\textrm{arcsinh}\sqrt{\frac{\tilde{B}_{0}-1}{2}}-\ln\left(2\tilde{B}_{0}\right).
\end{align}
As $\tilde{B}_{0}\rightarrow+\infty$, we find that 
\begin{equation}
\tilde{I}_{1}\simeq\ln\left(\frac{\tilde{B}_{0}-1}{\tilde{B}_{0}}\right)\simeq0.
\end{equation}
On the other hand, the second integral take the form,
\begin{equation}
\sum_{\mathbf{q}}\left[\frac{C^{2}}{\hbar^{2}q^{2}/m_{X}+\varepsilon_{c}}-\frac{C^{2}}{\hbar^{2}q^{2}/m_{X}+2C+2\Omega^{2}/\left(\delta-\mu\right)}\right]=\frac{m_{X}C^{2}}{4\pi\hbar^{2}}\tilde{I}_{2}
\end{equation}
where the dimensionless integral $\tilde{I}_{2}$ is 
\begin{equation}
\tilde{I}_{2}=\ln\left(\frac{2\tilde{B}_{0}}{\varepsilon_{c}/C}\right).
\end{equation}
Therefore, the dimensionless integral $\tilde{I}=\tilde{I}_{1}+\tilde{I}_{2}$
is 
\begin{equation}
\tilde{I}=\ln\left(\frac{\tilde{B}_{0}-1}{\tilde{B}_{0}}\right)+\ln\left(\frac{2\tilde{B}_{0}}{\varepsilon_{c}/C}\right)=\ln\left[\frac{2\Omega^{2}/\left(\delta-\mu\right)}{\varepsilon_{c}}\right],
\end{equation}
and we obtain that
\begin{equation}
\sum_{\mathbf{q}}\left[E_{\mathbf{q}+}+E_{\mathbf{q}-}-A_{\mathbf{q}}-B_{\mathbf{q}}+\frac{C^{2}}{\frac{\hbar^{2}q^{2}}{m_{X}}+\varepsilon_{c}}\right]=\frac{m_{X}C^{2}}{4\pi\hbar^{2}}\ln\left[\frac{2\Omega^{2}/\left(\delta-\mu\right)}{\varepsilon_{c}}\right].\label{eq:IntegralPolaritons}
\end{equation}
We note that, the above integral has also been numerically evaluated
(in suitable dimensionless form) for a given mass ratio $m_{X}/m_{\textrm{ph}}$.
We find that our analytic expression in Eq. (\ref{eq:IntegralPolaritons})
is essentially exact for a realistic mass ratio $m_{X}/m_{\textrm{ph}}\sim10^{4}$.
We note also that, if the light-matter coupling $\Omega=0$, we would
have $\tilde{B}_{0}=1$. The dimensionless integrals are then $\tilde{I}_{1}=1/2-\ln2$
and $\tilde{I}_{2}=\ln[2\mu_{B}/\varepsilon_{c}]$, respectively.
Therefore, we find that $\tilde{I}=1/2+\ln[\mu_{B}/\varepsilon_{c}]$,
which is $1/2$ if we take $\varepsilon_{c}=\mu_{B}$. We then recover
Eq. (\ref{eq:IntegralExcitons}), as one may anticipate.

By substituting Eq. (\ref{eq:IntegralPolaritons}) into Eq. (\ref{eq:OmegaBog-1}),
we finally obtain,
\begin{equation}
\varOmega=-\frac{m_{X}}{8\pi\hbar^{2}}\left[\mu+\frac{\Omega^{2}}{\delta-\mu}\right]^{2}\ln\left[\frac{2}{e^{2\gamma}}\frac{\hbar^{2}\left(\delta-\mu\right)}{m_{X}a_{s}^{2}\Omega^{2}}\right].\label{eq:OmegaBog2}
\end{equation}
By expanding $\mu=E_{LP}+\mu_{B}$, for small $\mu_{B}$, we have,
\begin{align}
\mu+\frac{\Omega^{2}}{\delta-\mu} & \simeq\frac{\mu_{B}}{X_{LP}^{2}},\\
\frac{\Omega^{2}}{\delta-\mu} & \simeq\left|E_{LP}\right|.
\end{align}
Therefore, we arrive at,
\begin{equation}
\varOmega=-\frac{\mu_{B}^{2}}{X_{LP}^{4}}\left(\frac{m_{X}}{8\pi\hbar^{2}}\right)\ln\left[\frac{2}{e^{2\gamma}}\frac{\hbar^{2}}{m_{X}a_{s}^{2}\left|E_{LP}\right|}\right].
\end{equation}

\section{Tunability of the exciton-exciton $s$-wave scattering length}

Although the underlying interaction between electrons and holes in
semiconductor quantum wells or atomically thin transition-metal-dichalcogenides
(TMD) monolayers is of the Coulomb type, the effective interaction
between composite excitons $V_{XX}(r)$ could be described by a short-range
Lennard-Jones potential, i.e., 
\begin{equation}
V_{XX}\simeq W\left[\left(\frac{a_{*}}{r}\right)^{12}-\left(\frac{a_{*}}{r}\right)^{6}\right],
\end{equation}
with a strength $W$ and a length scale $a_{*}$ comparable to the
excitonic Bohr radius $a_{X}$. At low temperature, only the $s$-wave
channel is important and we then can use a single $s$-wave scattering
length $a_{s}$ to characterize the effective interaction. This was
illustrated by a recent Monte-Carlo simulation in three dimensions
with the $1/r$ Coulomb interaction \cite{Golomedov2017}. It was
found that the exciton-exciton $s$-wave scattering length is comparable
to the exciton Bohr radius, $a_{s}\sim a_{X}$. An exact solution
for the four-body problem with long-range interaction such as the
Coulomb interaction is extremely difficult and is not available. 

In real materials, the Coulomb-like interactions among electrons and
holes take the following screened potential form \cite{Cudazzo2011},
\begin{equation}
V_{C}^{\sigma\sigma'}\left(r\right)=\chi_{\sigma\sigma'}\frac{\pi e^{2}}{2\varepsilon_{s}r_{0}}\left[H_{0}\left(\frac{r}{r_{0}}\right)-Y_{0}\left(\frac{r}{r_{0}}\right)\right],\label{eq:VC}
\end{equation}
where $\chi_{\sigma\sigma'}=+1$ for $\sigma=\sigma'$ and $\chi_{\sigma\sigma'}=-1$
for $\sigma\neq\sigma'$, and the spin index $\sigma$ stands for
either electrons or holes, $\varepsilon_{s}$ is the dielectric constant
of the substrate surrounding the quantum well or TMD monolayer, $H_{0}(x)$
and $Y_{0}(x)$ are respectively the Struve and Neumann functions,
and $r_{0}$ is an effective screening length. This particular form
of the Coulomb-like interaction is due to the large difference in
the dielectric constants between the quantum well or TMD monolayer
and the substrate, which strongly modifies the Coulomb interaction
at short distance \cite{Cudazzo2011}. As a result, the exciton-exciton
$s$-wave scattering length $a_{s}$ could depend on the effective
screening length $r_{0}$ and the dielectric constant $\varepsilon_{s}$.
Therefore, by carefully designing/choosing the materials, we may have
the ability to tune the exciton-exciton $s$-wave scattering length
$a_{s}$.

\section{Universal Tan relations}

In 2005, Shina Tan derived a set of exact universal relations to describe
the short-range, large-momentum and high-energy behaviors of a quantum
many-body system interacting via a short-range potential \cite{Tan2008a,Tan2008b,Braaten2008}.
These relations are linked by Tan's contact coefficient $\mathcal{I}$.
In ultracold atomic physics, the universal Tan relations help a lot
for us to understand the fundamental interacting Fermi gases and Bose
gases. Here, we generalize Tan relations to the exciton-polariton
system, following the work by Braaten and Platter \cite{Braaten2008}.

For exciton-polaritons, the contact coefficient can be formally defined
by, 
\begin{equation}
\mathcal{I}=\frac{m_{X}^{2}g^{2}}{\hbar^{4}}\int d\mathbf{r}\left\langle X^{\dagger}(\mathbf{r})X^{\dagger}(\mathbf{r})X(\mathbf{r})X(\mathbf{r})\right\rangle ,
\end{equation}
where the average $\left\langle ...\right\rangle $ is taken for \emph{any}
quantum states. It is worth noting that the bare exciton-exciton interaction
strength $g$ is vanishingly small in the sense of its regularization,
see Eq. (\ref{eq:gReg}). However, this smallness will be compensated
by the divergence in $\left\langle X^{\dagger}X^{\dagger}XX\right\rangle $,
resulting in a finite contact coefficient. To see this, let us recall
that
\begin{equation}
\frac{\partial g}{\partial\ln a_{s}}=-g^{2}\left(\frac{\partial g^{-1}}{\partial\ln a_{s}}\right)=\frac{m_{X}}{2\pi\hbar^{2}}g^{2},
\end{equation}
and apply the Hellmann--Feynman theorem to the total energy of the
system,
\begin{equation}
\left(\frac{\partial E}{\partial\ln a_{s}}\right)_{S,N}=\left\langle \frac{\partial\mathscr{H}}{\partial\ln a_{s}}\right\rangle =\frac{1}{2}\left(\frac{\partial g}{\partial\ln a_{s}}\right)\int d\mathbf{r}\left\langle X^{\dagger}(\mathbf{r})X^{\dagger}(\mathbf{r})X(\mathbf{r})X(\mathbf{r})\right\rangle =\frac{m_{X}}{4\pi\hbar^{2}}\frac{\hbar^{4}}{m_{X}^{2}}\mathcal{I},
\end{equation}
where the subscripts ``$S$'' and ``$N$'' indicate that the change
of the energy is taken under \emph{adiabatic} condition at a given
number of particles. Therefore, we obtain the adiabatic relation,
\begin{equation}
\left(\frac{\partial E}{\partial\ln a_{s}}\right)_{S,N}=\frac{\hbar^{2}}{4\pi m_{X}}\mathcal{I}.
\end{equation}
If we consider the grand-canonical ensemble, where the chemical potential
is fixed, by using standard thermodynamic relations, we can re-cast
Tan's adiabatic relation into the form,
\begin{equation}
\left(\frac{\partial\varOmega}{\partial\ln a_{s}}\right)_{S,\mu}=\frac{\hbar^{2}}{4\pi m_{X}}\mathcal{I}.\label{eq:AdibaticRelation-1}
\end{equation}
By using the thermodynamic potential within the Bogoliubov approximation,
i.e., Eq. (\ref{eq:OmegaBog2}), we immediately obtain the contact
coefficient predicted by the Bogoliubov theory:
\begin{equation}
\mathcal{I}=\frac{m_{X}^{2}}{\hbar^{4}}\left[\mu+\frac{\Omega^{2}}{\delta-\mu}\right]^{2}.\label{eq:ContactCoefficient}
\end{equation}

Let us now examine the kinetic energy $\mathscr{T}_{X}$ and interaction
energy $\mathscr{E}_{\textrm{int}}$ of excitons,
\begin{align}
\mathscr{T}_{X}+\mathscr{E}_{\textrm{int}} & =\sum_{\mathbf{q}}\frac{\hbar^{2}q^{2}}{2m_{X}}n_{X}\left(q\right)+\frac{g^{2}}{2g}\int d\mathbf{r}\left\langle X^{\dagger}(\mathbf{r})X^{\dagger}(\mathbf{r})X(\mathbf{r})X(\mathbf{r})\right\rangle ,\\
 & =\sum_{\mathbf{q}}\frac{\hbar^{2}q^{2}}{2m_{X}}n_{X}\left(q\right)+\left\{ \frac{m_{X}}{8\pi\hbar^{2}}\ln\left[\frac{4}{e^{2\gamma}}\frac{\hbar^{2}}{m_{X}a_{s}^{2}\varepsilon_{c}}\right]-\frac{1}{2}\sum_{\mathbf{q}}\left[\frac{\hbar^{2}\mathbf{q}^{2}}{m_{X}}+\varepsilon_{c}\right]^{-1}\right\} \frac{\hbar^{4}}{m_{X}^{2}}\mathcal{I}.
\end{align}
We may take the infrared cut-off energy $\varepsilon_{c}=\hbar^{2}/(m_{X}a_{s}^{2})$
to simplify the equation. This leads to Tan's energy relation,

\begin{equation}
\mathscr{T}_{X}+\mathscr{E}_{\textrm{int}}=\sum_{\mathbf{q}}\frac{\hbar^{2}q^{2}}{2m_{X}}\left[n_{X}(q)-\frac{\mathcal{I}}{q^{2}\left(q^{2}+a_{s}^{-2}\right)}\right]+\frac{\left(\ln2-\gamma\right)\hbar^{2}\mathcal{I}}{4\pi m_{X}},\label{eq:EnergyRelation-1}
\end{equation}
It is clear from the energy relation that the excitonic momentum distribution
must have a universal $q^{-4}$ tail:
\begin{equation}
n_{X}\left(q\rightarrow\infty\right)=\frac{\mathcal{I}}{q^{2}\left(q^{2}+a_{s}^{-2}\right)}\simeq\frac{\mathcal{I}}{q^{4}}.
\end{equation}

\subsection{The momentum distribution of photons and excitons}

One may wonder that the photonic momentum distribution $n_{\textrm{ph}}(q)$
may similarly develop a universal $q^{-4}$ tail, as naively anticipated
from the scenario of polariton quasiparticles. However, as we examine
directly in the following, this is not the case. The absence of a
universal tail in $n_{\textrm{ph}}(q)$ is understandable, since it
is a large-momentum, high-energy behavior, which can not be captured
by the low-energy quasiparticle picture.

To see this, let us calculate the momentum distribution of photons
and excitons within the Bogoliubov theory. The Green function is given
by, 
\begin{equation}
\mathscr{D}\left(\mathbf{q},i\nu_{n}\right)=\left[\begin{array}{cccc}
i\nu_{n}-A_{\mathbf{q}} & 0 & -\Omega & 0\\
0 & -i\nu_{n}-A_{\mathbf{q}} & 0 & -\Omega\\
-\Omega & 0 & i\nu_{n}-B_{\mathbf{q}} & -C\\
0 & -\Omega & -C & -i\nu_{n}-B_{\mathbf{q}}
\end{array}\right]^{-1}.
\end{equation}
By taking the inverse of the above four by four matrix, we find that
the Green function of photons,
\begin{equation}
\mathscr{G}_{\textrm{ph}}\left(\mathbf{q},i\nu_{n}\right)=\mathscr{D}_{11}\left(\mathbf{q},i\nu_{n}\right)=\frac{\left(i\nu_{n}\right)^{3}+A_{\mathbf{q}}\left(i\nu_{n}\right)^{3}-(B_{\mathbf{q}}^{2}-C^{2}+\Omega^{2})i\nu_{n}-\left[A_{\mathbf{q}}\left(B_{\mathbf{q}}^{2}-C^{2}\right)-B_{\mathbf{q}}\Omega^{2}\right]}{\left[\left(i\nu_{n}\right)^{2}-E_{\mathbf{q}+}^{2}\right]\left[\left(i\nu_{n}\right)^{2}-E_{\mathbf{q}-}^{2}\right]}.
\end{equation}
Integrating over the bosonic Matsubara frequencies $i\nu_{n}$, we
obtain,
\begin{equation}
n_{\textrm{ph}}\left(\mathbf{q}\right)=\frac{1}{2}\left[\frac{A_{\mathbf{q}}}{E_{\mathbf{q}+}+E_{\mathbf{q}-}}+\frac{A_{\mathbf{q}}\left(B_{\mathbf{q}}^{2}-C^{2}\right)-B_{\mathbf{q}}\Omega^{2}}{\left(E_{\mathbf{q}+}+E_{\mathbf{q}-}\right)E_{\mathbf{q}+}E_{\mathbf{q}-}}-1\right].
\end{equation}
At large momentum, both $A_{\mathbf{q}}$ and $B_{\mathbf{q}}$ are
much larger than $C$ and $\Omega$. We may use Eq. (\ref{eq:EqpApprox})
and Eq. (\ref{eq:EqmApprox}) to perturbatively expand $E\mathbf{_{q\pm}}$.
Thus, we find that, when $q\rightarrow\infty$, 
\begin{equation}
n_{\textrm{ph}}\left(\mathbf{q}\right)=\frac{C^{2}}{2\left(A_{\mathbf{q}}+B_{\mathbf{q}}\right)}\left[\frac{A_{\mathbf{q}}}{2\left(A_{\mathbf{q}}+B_{\mathbf{q}}\right)B_{\mathbf{q}}}-\frac{1}{B_{\mathbf{q}}}+\frac{1}{2\left(A_{\mathbf{q}}+B_{\mathbf{q}}\right)}+\frac{1}{2B_{\mathbf{q}}}+\mathcal{O}\left(q^{-4}\right)\right]=\mathcal{O}\left(q^{-6}\right).
\end{equation}
Therefore, we conclude that within the Bogoliubov theory, there is
no $q^{-4}$ tail in the photonic momentum distribution.

For the excitonic momentum distribution, the Green function of excitons
takes the form,
\begin{equation}
\mathscr{G}_{X}\left(\mathbf{q},i\nu_{n}\right)=\mathscr{D}_{33}\left(\mathbf{q},i\nu_{n}\right)=\frac{\left(i\nu_{n}\right)^{3}+B_{\mathbf{q}}\left(i\nu_{n}\right)^{3}-(A_{\mathbf{q}}^{2}+\Omega^{2})i\nu_{n}-A_{\mathbf{q}}\left(A_{\mathbf{q}}B_{\mathbf{q}}-\Omega^{2}\right)}{\left[\left(i\nu_{n}\right)^{2}-E_{\mathbf{q}+}^{2}\right]\left[\left(i\nu_{n}\right)^{2}-E_{\mathbf{q}-}^{2}\right]},
\end{equation}
and the momentum distribution is,
\begin{equation}
n_{X}\left(\mathbf{q}\right)=\frac{1}{2}\left[\frac{B_{\mathbf{q}}}{E_{\mathbf{q}+}+E_{\mathbf{q}-}}+\frac{A_{\mathbf{q}}\left(A_{\mathbf{q}}B_{\mathbf{q}}-\Omega^{2}\right)}{\left(E_{\mathbf{q}+}+E_{\mathbf{q}-}\right)E_{\mathbf{q}+}E_{\mathbf{q}-}}-1\right].
\end{equation}
Let us similarly express $E\mathbf{_{q\pm}}$ in terms of $A_{\mathbf{q}}$
and $B_{\mathbf{q}}$ in the large momentum limit. We obtain, for
$q\rightarrow\infty$,
\begin{equation}
n_{X}\left(\mathbf{q}\right)=\frac{C^{2}}{4\left(A_{\mathbf{q}}+B_{\mathbf{q}}\right)^{2}}\left[1+\frac{A_{\mathbf{q}}}{B_{\mathbf{q}}}+\frac{A_{\mathbf{q}}\left(A_{\mathbf{q}}+B_{\mathbf{q}}\right)}{B_{\mathbf{q}}^{2}}+\mathcal{O}\left(q^{-2}\right)\right]=\frac{C^{2}}{4B_{\mathbf{q}}^{2}}+\mathcal{O}\left(q^{-6}\right)\simeq\frac{m_{X}^{2}}{\hbar^{4}}\left[\mu+\frac{\Omega^{2}}{\delta-\mu}\right]^{2}q^{-4}.
\end{equation}
Therefore, the contact coefficient extracted from the tail of $n_{X}(\mathbf{q})$
is the same as that calculated using the adiabatic relation, see Eq.
(\ref{eq:ContactCoefficient}).

\section{Multiple quantum wells}

In semiconductor quantum wells, such as GaAs, multiple quantum wells
are used to \emph{enhance} the light-matter coupling \cite{Estrecho2019}.
Here, we show that the same results of the polariton-polariton interaction
strength and Tan contact coefficient can be derived, up to a trivial
factor of $N$, where $N$ is the number of quantum wells.

In the presence of $N$ quantum wells, the bosonic model Hamiltonian
$\mathscr{H}=\mathscr{H}_{0}+\mathscr{H}_{\textrm{LM}}+\mathscr{H}_{\textrm{int}}$
takes the form, 
\begin{eqnarray}
\mathscr{H}_{0} & = & \sum_{\mathbf{q}}\left(\frac{\hbar^{2}\mathbf{q}^{2}}{2m_{\textrm{ph}}}+\delta-\mu\right)\phi_{\mathbf{q}}^{\dagger}\phi_{\mathbf{q}}+\sum_{i=1}^{N}\sum_{\mathbf{q}}\xi_{\mathbf{q}}X_{i\mathbf{q}}^{\dagger}X_{i\mathbf{q}},\\
\mathscr{H}_{\textrm{LM}} & = & \frac{\Omega}{\sqrt{N\mathcal{S}}}\sum_{i=1}^{N}\sum_{\mathbf{q}}\left[\phi_{\mathbf{q}}^{\dagger}X_{i\mathbf{q}}+X_{i\mathbf{q}}^{\dagger}\phi_{\mathbf{q}}\right],\\
\mathscr{H}_{\textrm{int}} & = & \frac{g}{2\mathcal{S}}\sum_{i=1}^{N}\sum_{\mathbf{q}\mathbf{q}'\mathbf{k}}X_{i\frac{\mathbf{k}}{2}+\mathbf{q}}^{\dagger}X_{i\frac{\mathbf{k}}{2}-\mathbf{q}}^{\dagger}X_{i\frac{\mathbf{k}}{2}-\mathbf{q}'}X_{i\frac{\mathbf{k}}{2}+\mathbf{q}'}.\label{eq: HamiCoul-1}
\end{eqnarray}
Here, $i=1,...,N$ is the index of the quantum wells. Each quantum
well is assumed to be identical and couples to the cavity with the
same light-matter coupling strength $\Omega/\sqrt{N}$.

As before, we assume that photon field and exciton fields condensate
at the zero-momentum states with condensate wave-functions $\phi_{0}$
and $X_{0}$. At the mean-field level, the thermodynamic potential
is
\begin{equation}
\varOmega_{0}\left(\mu\right)=\left(\delta-\mu\right)\phi_{0}^{2}+2\Omega\phi_{0}\tilde{X}_{0}-\mu\tilde{X}_{0}^{2}+\frac{g}{2N}\tilde{X}_{0}^{4},
\end{equation}
which takes the same form as in the case of single quantum well, after
we introduce $\tilde{X}_{0}^{2}\equiv NX_{0}^{2}$. By minimizing
the mean-field thermodynamic potential with respect to $\phi_{0}$
and $\tilde{X}_{0}$, we obtain,
\begin{equation}
gX_{0}^{2}=\frac{g}{N}\tilde{X}_{0}^{2}=\mu+\frac{\Omega^{2}}{\delta-\mu},
\end{equation}
and 
\begin{equation}
\varOmega_{0}=-N\frac{1}{2g}\left[\mu+\frac{\Omega^{2}}{\delta-\mu}\right]^{2}.
\end{equation}

Beyond mean-field, we keep the bilinear terms in the field operators
and obtain the Bogoliubov action,
\begin{equation}
\mathscr{H}_{\textrm{Bog}}=\sum_{\mathcal{Q}=\left(\mathbf{q},i\nu_{n}\right)}\left[\delta\phi_{\mathcal{Q}}^{\dagger},\delta\phi_{-\mathcal{Q}},\delta X_{1,\mathcal{Q}}^{\dagger},\cdots,\delta X_{N,\mathcal{Q}}^{\dagger},\delta X_{1,-\mathcal{Q}},\cdots,\delta X_{N,-\mathcal{Q}}\right]\left[-\mathscr{D}^{-1}\left(\mathcal{Q}\right)\right]\left[\begin{array}{c}
\delta\phi_{\mathcal{Q}}\\
\delta\phi_{-\mathcal{Q}}^{\dagger}\\
\delta X_{1,\mathcal{Q}}\\
\vdots\\
\delta X_{N,\mathcal{Q}}\\
\delta X_{1,-\mathcal{Q}}^{\dagger}\\
\vdots\\
\delta X_{N,-\mathcal{Q}}^{\dagger}
\end{array}\right],
\end{equation}
where the inverse Green function,

\begin{equation}
-\mathscr{D}^{-1}\left(\mathcal{Q}\right)=\left[\begin{array}{cccccccc}
-i\nu_{n}+A_{\mathbf{q}} & 0 & \frac{\Omega}{\sqrt{N}} & \cdots & \frac{\Omega}{\sqrt{N}} & 0 & \cdots & 0\\
0 & i\nu_{n}+A_{\mathbf{q}} & 0 & \cdots & 0 & \frac{\Omega}{\sqrt{N}} & \cdots & \frac{\Omega}{\sqrt{N}}\\
\frac{\Omega}{\sqrt{N}} & 0 & -i\nu_{n}+B_{\mathbf{q}} & 0 & 0 & C & 0 & 0\\
\vdots & \vdots & 0 & \cdots & 0 & 0 & \cdots & 0\\
\frac{\Omega}{\sqrt{N}} & 0 & 0 & 0 & -i\nu_{n}+B_{\mathbf{q}} & 0 & 0 & C\\
0 & \frac{\Omega}{\sqrt{N}} & C & 0 & 0 & i\nu_{n}+B_{\mathbf{q}} & 0 & 0\\
\vdots & \vdots & 0 & \cdots & 0 & 0 & \cdots & 0\\
0 & \frac{\Omega}{\sqrt{N}} & 0 & 0 & C & 0 & 0 & i\nu_{n}+B_{\mathbf{q}}
\end{array}\right].
\end{equation}
By solving the poles of the Green function, we find that there is
$N-1$ degenerate eigenvalues 
\begin{equation}
E_{\mathbf{q}}=\sqrt{B_{\mathbf{q}}^{2}-C^{2}},
\end{equation}
in addition to the eigenvalues $E_{\mathbf{q}+}$ and $E_{\mathbf{q}-}$.
Therefore, the fluctuation thermodynamic potential is given by,

\begin{equation}
\delta\varOmega_{g}^{(T=0)}=\frac{1}{2}\sum_{\mathbf{q}}\left[E_{\mathbf{q}+}+E_{\mathbf{q}-}+\left(N-1\right)\sqrt{B_{\mathbf{q}}^{2}-C^{2}}-A_{\mathbf{q}}-NB_{\mathbf{q}}\right].
\end{equation}
By adding the two thermodynamic potentials and removing the bare interaction
strength $g$, we obtain,

\begin{equation}
\varOmega=-N\frac{m_{X}C^{2}}{8\pi\hbar^{2}}\ln\left[\frac{4}{e^{2\gamma}}\frac{\hbar^{2}}{m_{X}a_{s}^{2}\varepsilon_{c}}\right]+\frac{1}{2}\sum_{\mathbf{q}}\left[E_{\mathbf{q}+}+E_{\mathbf{q}-}+\left(N-1\right)\sqrt{B_{\mathbf{q}}^{2}-C^{2}}-A_{\mathbf{q}}-NB_{\mathbf{q}}+\frac{NC^{2}}{\hbar^{2}q^{2}/m_{X}+\varepsilon_{c}}\right].\label{eq:}
\end{equation}
By repeating the steps in Appendix A, in the limit of an infinite
mass ratio, it is easy to see that,
\begin{equation}
\sum_{\mathbf{q}}\left[E_{\mathbf{q}+}+E_{\mathbf{q}-}+\left(N-1\right)\sqrt{B_{\mathbf{q}}^{2}-C^{2}}-A_{\mathbf{q}}-NB_{\mathbf{q}}+\frac{NC^{2}}{\hbar^{2}q^{2}/m_{X}+\varepsilon_{c}}\right]=N\frac{m_{X}C^{2}}{4\pi\hbar^{2}}\ln\left[\frac{2\Omega^{2}/\left(\delta-\mu\right)}{\varepsilon_{c}}\right].
\end{equation}
Therefore, we obtain
\begin{equation}
\varOmega=-N\frac{m_{X}}{8\pi\hbar^{2}}\left[\mu+\frac{\Omega^{2}}{\delta-\mu}\right]^{2}\ln\left[\frac{2}{e^{2\gamma}}\frac{\hbar^{2}\left(\delta-\mu\right)}{m_{X}a_{s}^{2}\Omega^{2}}\right].\label{eq:OmegaBog3}
\end{equation}
It is readily seen that the thermodynamic potential is trivially enlarged
by a factor of $N$, in the case of $N$ quantum wells. As a result,
the density is enlarged by $N$ times at a given chemical potential
$\mu$ and hence the polariton-polariton interaction strength is \emph{reduced}
by a factor of $N$, i.e., 
\begin{equation}
g_{PP}=\frac{X_{LP}^{4}}{N}\left(\frac{4\pi\hbar^{2}}{m_{X}}\right)\ln^{-1}\left[\frac{2}{e^{2\gamma}}\frac{\hbar^{2}}{m_{X}a_{s}^{2}\left|E_{LP}\right|}\right].\label{eq:gPPBog-1}
\end{equation}
In line with this factor of $N$ scaling, Tan contact coefficient
within the Bogoliubov theory is now given by,
\begin{equation}
\mathcal{I}=\frac{1}{N^{2}}\frac{\left(16\pi^{2}n^{2}\right)X_{LP}^{4}}{\ln^{2}\left[e^{2\gamma}m_{X}a_{s}^{2}\left|E_{LP}\right|/\left(2\hbar^{2}\right)\right]},
\end{equation}
which is reduced by a factor of $1/N^{2}$.

\end{widetext}

\begin{thebibliography}{10}
\bibitem{NoteBornApproximation}In scattering theory, the Born approximation
amounts to truncating the Lippmann-Schwinger equation to the first
order and taking the bare scattering potential to replace the two-body
$T$-matrix. In quantum many-body approach, the Born approximation
result for the inter-particle interaction strength can often be conveniently
obtained, by taking mean-field approximation to the many-body theory.
In this context, we take the Born approximation and mean-field approximation
as exchangeable terminology.
\end{thebibliography}
\end{document}